\documentclass{sig-alternate}
\usepackage[numbers]{natbib}
\usepackage{hyperref}
\usepackage{tikz}
\usepackage{cpc_header}
\pdfoutput=1

\hypersetup{
  pdftitle={2HOT: An Improved Parallel Hashed Oct-Tree\\N-Body Algorithm for Cosmological Simulation},
  pdfauthor={Michael S. Warren, Theoretical Division, LANL},
  colorlinks=false
}

\begin{document}

\title{2HOT: An Improved Parallel Hashed Oct-Tree\\N-Body Algorithm for Cosmological Simulation\\
{\small\sl Best Paper Finalist, SC '13.}\\[-.25in]}
\numberofauthors{1}
\author{
\alignauthor
Michael S. Warren\\
       \affaddr{Theoretical Division}\\
       \affaddr{Los Alamos National Laboratory}\\
       \email{msw@lanl.gov}
}
\toappear{
Permission to make digital or hard copies of part or all of this work for personal or classroom use is granted without fee provided that copies are not made or distributed for profit or commercial advantage and that copies bear this notice and the full citation on the first page. Copyrights for third-party components of this work must be honored. For all other uses, contact the Owner/Author.\\[.03in]
Copyright is held by the owner/author(s).\\[.03in]
{\confname{SC '13}} Nov 17-21 2013, Denver, CO, USA\\
ACM 978-1-4503-2378-9/13/11.\\
http://dx.doi.org/10.1145/2503210.2503220
}

\maketitle
\begin{abstract}
We report on improvements made over the past two decades to our
adaptive treecode N-body method (HOT).  A mathematical and
computational approach to the cosmological N-body problem is
described, with performance and scalability measured up to 256k
($2^{18}$) processors.  We present error analysis and scientific
application results from a series of more than ten 69 billion
($4096^3$) particle cosmological simulations, accounting for $4 \times
10^{20}$ floating point operations.  These results include the first
simulations using the new constraints on the standard model of
cosmology from the Planck satellite.  Our simulations set a new
standard for accuracy and scientific throughput, while meeting or
exceeding the computational efficiency of the latest generation of
hybrid TreePM N-body methods.
\end{abstract}

\terms{Computational Cosmology, N-body, Fast Multipole Method}

\section{Introduction}

We first reported on our parallel N-body algorithm (HOT) 20 years
ago~\cite{warren93} (hereafter WS93).  Over the same timescale,
cosmology has been transformed from a qualitative to a quantitative
science.  Constrained by a diverse suite of
observations~\cite{smoot92,spergel03,tegmark04,riess04,planckcollaboration13},
the parameters describing the large-scale Universe are now known to
near 1\% precision.  In this paper, we describe an improved version of
our code (2HOT), and present a suite of simulations which probe the
finest details of our current understanding of cosmology.

\begin{figure}[h!]
\resizebox{1.08\columnwidth}{!}{\includegraphics{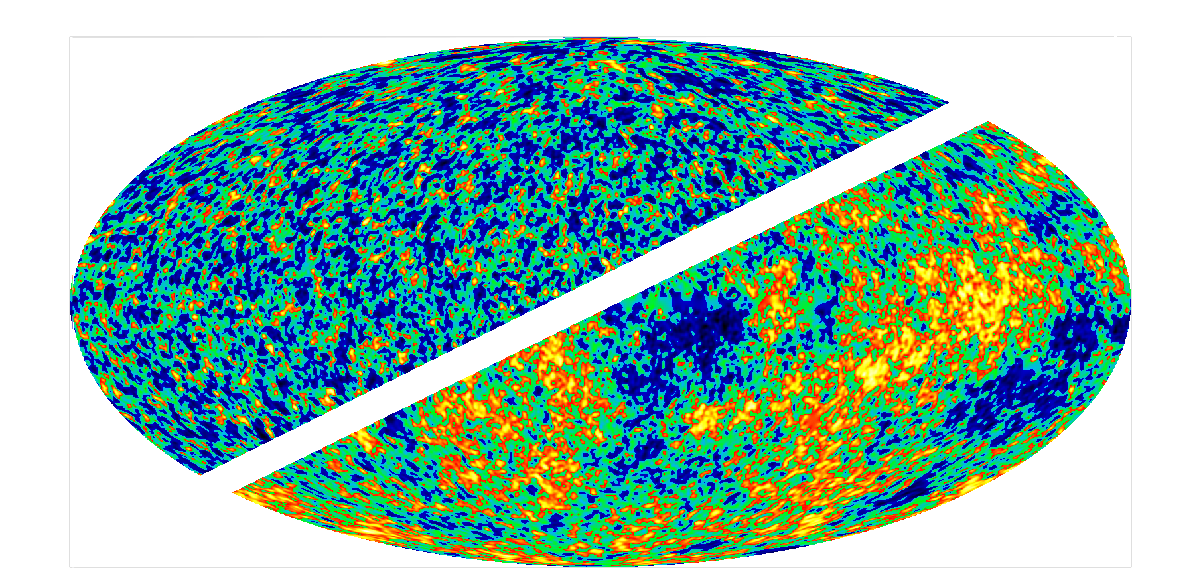}}
\resizebox{1.08\columnwidth}{!}{\includegraphics{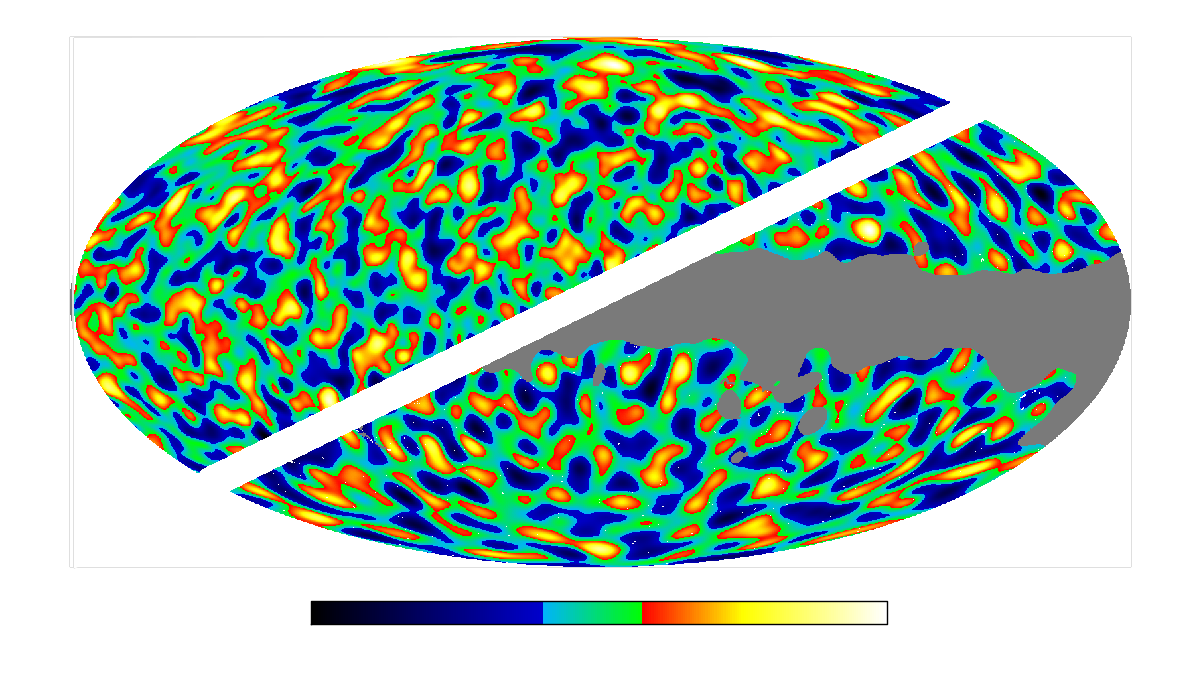}}
\caption{Recent results from the Planck
  satellite~\protect\cite{planckcollaboration13} compared with
  light-cone output from 2HOT.  We present our numerical simulation
  results in the same
  HEALPix\protect\footnotemark[1]~\protect\cite{gorski05} Mollewide
  projection of the celestial sphere used by Planck.  The upper figure
  shows the density of dark matter in a 69 billion particle simulation
  (upper left) compared with the fluctuations in the cosmic microwave
  background.  The obvious difference in the upper panel is due to the
  imperfect removal of sources within our galaxy in the Planck data;
  the statistical measurements of the smaller details match precisely
  between the observation and simulation. The lower figure shows the
  simulation compared with the gravitational lensing signal measured
  by Planck.}\ \\[-.25in]
\label{fig:Planck}
\end{figure}

Computer simulations enable discovery.  In the words of the Astronomy
and Astrophysics Decadal Survey, ``Through computer modeling, we
understand the deep implications of our very detailed observational
data and formulate new theories to stimulate further
observations''~\cite{nationalresearchcouncil10}.  The only way to
accurately model the evolution of dark matter in the Universe is
through the use of advanced algorithms on massively parallel computers
(see \cite{kuhlen12} for a recent review).  The origin of cosmic
structure and the global evolution of the Universe can be probed by
selecting a set of cosmological parameters, modeling the growth of
structure, and then comparing the model to the observations 
(Figure~\ref{fig:Planck}).
\footnotetext[1]{\url{http://healpix.jpl.nasa.gov}}

Computer simulations are playing an increasingly important role in the
modern scientific method, yet the exponential pace of growth in the
size of calculations does not
necessarily translate into better tests of our scientific models or
increased understanding of our Universe.  Anywhere the relatively slow
growth in the capacity of human attention intersects with the
exponential explosion of information, new tensions are created.  The
timespan between the completion of a large simulation and the publication
of scientific results based upon it is now often a year or more, and is
growing longer instead of shorter.  In the application described here,
the sheer complexity of managing the volume of information in many
layers of data and code has required additional software tools to be
developed.  We have written substantially more lines of software for
data analysis, generating initial conditions, testing and task
management than are present in the 2HOT code base. The scale of
simulations requires most of these ancillary tools to be parallel
as well.

High-performance computing (HPC) allows us to probe more questions
with increased resolution and reduced statistical uncertainty, leading
to new scientific discoveries.  However, reducing the statistical
errors more often than not uncovers systematic errors previously
masked by statistical variance.  Addressing these details takes us out
of realm of HPC into applied mathematics, software engineering and
data analysis. However, without progress on all fronts, the
over-arching scientific questions can not be answered.  A corollary of
this point is that making a code faster is often a poor investment
when the aim is to answer a particular scientific question.  More
important than speed is the code's applicability to the problem,
correctness, and even less tangible properties such as robustness and
maintainability.  For those reasons, we focus here on the wide variety
of changes made to 2HOT over the past two decades which have enabled
us to produce the state-of-the-art scientific results presented in
Section~\ref{sec:science}.

One of our first scientific N-body simulations of dark
matter~\cite{warren91a} used 1.1 million particles and was performed
on the 64-node Caltech/JPL Mark III hypercube in 1990.  The simulation
was completed in 60 hours, sustaining 160 Mflop/s with a parallel
efficiency of 85\%.  In 2012 we used 2HOT on 262 thousand processors
with over one trillion ($10^{12}$) particles, sustaining in excess of
1.6 Petaflops with a parallel efficiency of 90\%~\cite{warren12}.
Since our first parallel treecode simulations, the message-passing
programming model, time to solution and parallel efficiency are nearly
the same, but the problem size has increased by a factor of a million,
and performance a factor of 10 million.  

Since WS93, HOT was been extended and optimized to be applicable to
more general problems such as incompressible fluid flow with the
vortex particle method~\cite{ploumhans02} and astrophysical gas
dynamics with smoothed particle
hydrodynamics~\cite{fryer02a,fryer06,ellinger12}.  The code also won
the Gordon Bell performance prize and price/performance prize in
1997~\cite{warren97a} and 1998~\cite{warren98}. It was an early
driver of Linux-based cluster architectures~\cite{warren97a,warren97b,warren03} and helped
call attention to power issues~\cite{warren02a,feng03}. Perhaps surprisingly (given
that WS93 was presented at the same conference as the draft MPI 1.0
standard), the fundamental HPC abstractions in the code have changed
little over two decades, while more significant changes have been
required in its mathematical and cosmological
underpinnings.

\section{Mathematical Approach}

\subsection{Equations of Motion}

The mathematical equations governing the evolution of structure in an
expanding Universe are generally solved using comoving coordinates, $
\vec{x} = \vec{r}/a(t) $.  $\vec{r}$ is the ``proper'' coordinate,
while the scale factor $a(t)$ is defined via the Friedmann equation
\be
\label{eq:scale}
(H/H_0)^2 = \Omega_R/a^4 + \Omega_M/a^3 + \Omega_k/a^2 + \Omega_{DE}
\ee
in terms of the Hubble parameter $H \equiv \dot{a}/a$
and the densities of the various components of the Universe; radiation
in the form of photons and ultra-relativistic particles ($\Omega_R$),
mass in the form of cold dark matter and baryons ($\Omega_M$), spatial
curvature ($\Omega_k$) and dark energy or a cosmological constant
($\Omega_{DE}$).  The particle dynamics are defined in terms of the
motion relative to the background model, the scale factor and the
acceleration due to gravity~\cite{peebles80},
\be
\label{eq:eom}
{d\vec{v_i} \over dt} + 2 {\dot{a} \over a} \vec{v_i}  = - {1 \over a^3} \sum_{i \neq j} {G m_j \vec{x}_{ij} \over \abs{x_{ij}}^3}
\ee
Cosmological evolution codes most often account
for cold dark matter, baryons and dark energy.  The Boltzmann solvers
which calculate the power spectrum of density perturbations use all of
the components, including photons and massless and massive neutrinos.
For precise computations, it is now necessary to include these other
species.  Using the parameters of the Planck 2013 cosmological model,
the age of the Universe is 3.7 million years older if photons and
radiation from massless neutrinos are not treated correctly.  The linear
growth factor from redshift 99 (an expansion of 100) changes by almost
5\% (from 82.8 to 79.0) under the same circumstances.  2HOT integrates
directly with the computation of the background quantities and growth
function provided by CLASS~\cite{lesgourgues11}, either in tabular
form or by linking directly with the CLASS library, and thereby
supports any cosmology which can be defined in CLASS.  2HOT
additionally maintains the ability to calculate the scale factor and
linear growth factor analytically (when radiation or non-trivial dark
energy is not included) in order to be able to directly compare with
codes which do not yet support them.

\subsection{Multipole Methods}

Using $N$ particles to represent the Universe, treecodes and fast
multipole methods reduce the $N^2$ scaling of the right-hand side of
equation \eqref{eq:eom} to $O(N)$ or $O(N \log N)$---a significant
savings for current cosmological simulations which use $N$ in the
range of $10^{10}$ to $10^{12}$.

\subsubsection{Background Subtraction}
\label{sec:bs}

Large cosmological simulations present a unique set of challenges for
multipole methods.  The Universe is nearly uniform at large scales.
This means the resultant acceleration on a particle from distant regions is a
sum of large terms which mostly cancel.  We can precisely quantify this effect
by looking at the variance of density in spheres of radius $r$, which is an
integral of the power spectrum convolved with a top-hat window,
\be
\int_0^\infty (dk/k) \delta_k^2 W(kr)^2
\ee
For a sphere of radius
100 Mpc/h, the variance is 0.068 of the mean value for the standard
model.  This value scales with the growth of cosmic structure over
time, so at the beginning of a simulation it will be a factor of
50-100 lower.  At early times when we calculate the acceleration
from a 100 Mpc cell in one direction, 99\% of that value will cancel
with a cell in the opposite direction, leaving a small remainder (the
``peculiar'' acceleration). This implies that the error tolerance
needed for these large cells is 100 times stricter than
for the short-range interactions.  For larger volumes or earlier
starting times, even more accuracy is required.  This suggests that
eliminating the background contribution from the partial acceleration
terms would be beneficial.

The mathematical equations describing the evolving Universe subtract
the uniform background, accounting for it in the evolution of the
scale factor $a(t)$.  Fourier-based codes do this automatically, since
the DC component has no dynamical effect.  For treecodes, the proper
approach is less obvious.  Essentially, we wish to convert the
always-positive mass distribution into density perturbations $\delta
\rho / \rho$.  These density contrasts can be positive or negative,
making the gravitational problem analogous to an electrostatics
problem, with positive and negative charges.

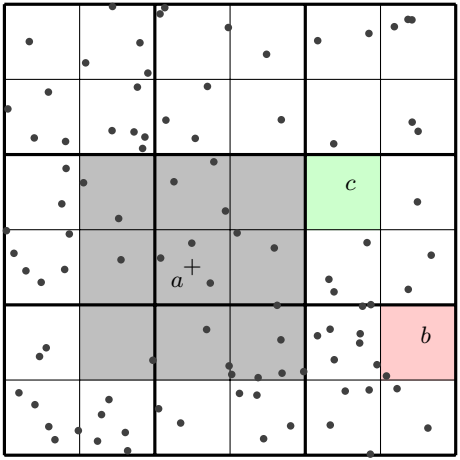
\begin{figure}
\centering
\begin{tikzpicture}
\filldraw [lightgray] (1,1) rectangle +(3,3);
\filldraw [green!20!white] (4,3) rectangle +(1,1);
\filldraw [red!20!white] (5,1) rectangle +(1,1);
\draw [step=1.0cm] (0,0) grid (6,6);
\draw [step=2.0cm,very thick] (0,0) grid (6,6);
\draw (2.5,2.5) node {+};
\draw (2.3,2.3) node {$a$};
\draw (5.6,1.6) node {$b$};
\draw (4.6,3.6) node {$c$};
\foreach \c in {
(4.385,1.270),
(3.094,2.957),
(4.723,1.492),
(1.519,3.149),
(3.676,1.535),
(5.418,5.792),
(4.872,2.003),
(4.821,2.828),
(1.291,0.540),
(2.491,2.821),
(0.027,2.987),
(3.485,5.334),
(3.979,1.112),
(1.080,5.220),
(2.069,5.871),
(5.220,0.885),
(4.165,5.515),
(3.626,1.994),
(5.369,2.206),
(4.314,2.340),
(0.489,2.300),
(2.254,3.638),
(5.491,3.370),
(4.378,4.143),
(2.077,2.623),
(1.390,0.740),
(0.820,3.815),
(0.585,4.831),
(5.184,5.701),
(4.382,2.173),
(2.737,2.290),
(0.763,3.344),
(2.784,3.903),
(3.805,0.390),
(0.467,1.313),
(4.531,0.853),
(2.537,4.215),
(5.630,0.361),
(1.55,2.6),
(1.974,1.263),
(2.147,4.458),
(4.953,1.204),
(4.761,1.982),
(1.868,4.233),
(3.373,1.032),
(5.421,4.431),
(2.688,1.672),
(0.397,4.221),
(4.162,1.589),
(0.863,2.943),
(1.638,0.059),
(4.846,5.611),
(0.983,0.327),
(4.332,0.401),
(1.907,5.084),
(5.079,1.054),
(2.987,1.188),
(2.343,0.428),
(0.556,1.429),
(0.813,4.174),
(1.431,4.318),
(2.699,4.906),
(0.331,5.503),
(0.590,0.380),
(0.672,0.206),
(4.330,1.676),
(1.802,5.487),
(1.837,4.081),
(1.607,0.302),
(5.500,4.309),
(4.730,1.616),
(2.132,5.954),
(2.977,5.690),
(0.801,2.471),
(0.286,2.453),
(1.052,3.625),
(3.022,1.074),
(2.050,0.618),
(3.691,1.091),
(4.865,0.014),
(3.681,4.463),
(0.046,4.607),
(1.238,0.187),
(1.768,4.897),
(3.588,2.757),
(5.365,5.798),
(2.939,3.249),
(1.723,4.301),
(1.437,5.970),
(3.125,0.822),
(3.446,0.219),
(5.674,2.661),
(0.193,0.830),
(0.406,0.672),
(0.127,2.686),
(3.357,0.799),
(4.849,0.868)
}
    \fill [black!75!white] \c circle (0.05);
\end{tikzpicture}
\caption{An illustration of background subtraction, which greatly
  improves the performance of the treecode algorithm for nearly
  uniform mass distributions (such as large-volume cosmological
  simulations, especially at early times). The bodies inside cell $a$
  interact with the bodies and cells inside the gray shaded area as usual.
  Bodies inside cell $a$ interact with all other cells ($b$, for example)
  after the background contribution of a uniform density cube is
  subtracted from the multipole expansion.  Empty cell $c$ (which
  would be ignored in the usual algorithm) must have its background
  contribution subtracted as well.  The background
  contribution of the gray shaded area to the calculated force and
  potential of the bodies in $a$ is removed analytically.}
\label{fig:bs}
\end{figure}

Since we wish to retain the particle-based representation of the
density, the background subtraction can be obtained by adding the
multipole expansion of a cube of uniform negative density to each
interaction.  Since the multipole expansion of a cube is fairly simple
due to symmetries, this can be done with a few operations if the
multipole expansions are with respect to the cell centers (rather than
the center of mass).  This in turn adds a few operations to the
interaction routines, since dipole moments are now present.  At scales
near the inter-particle separation, this approach breaks down, since
any empty cells which would be ignored in a direct summation must be
accounted for, as well as requiring high-order expansions for
neighboring cells with only a few particles, which would normally be
calculated with cheaper monopole interactions.  Rather than modify
each interaction for the near field, we define a larger cube which
approximately surrounds the local region of empty and single particle
cells and calculate the background acceleration within the surrounding
cell (Figure~\ref{fig:bs}).  This acceleration term can be done with a
multipole and local expansion, or our current approach of using the
analytic expression for the force inside a uniform
cube~\cite{waldvogel76,seidov00}.

A subtle point is that in the far-field we only want to subtract the
uniform background expansion up to the same order as the multipole
expansion of the matter to minimize the error.  If a cube of particles
is expanded to order $p=4$, the $p=6$ and higher multipoles from the
background are not included, so they should not be subtracted.  Using
background subtraction increases the cost of each interaction
somewhat, but results in a huge improvement in overall efficiency,
since many fewer interactions need to be computed.  At early times we
have measured an improvement of a factor of five.  The multipole
acceptance criterion (MAC) based on an absolute error also becomes
much better behaved, leading to improved error behavior as well.

\subsubsection{Multipole Error Bounds}
A critical ingredient of any optimized multipole method is the
mathematical machinery to bound or estimate the error in the
interactions.  The methods we previously developed
\cite{salmon94,warren95a} allow us to dynamically decide between using
different orders of expansion or refinement, automatically choosing the
most efficient method to achieve a given accuracy.

The expressions we derived in \cite{warren95a} support
methods which use both multipole and local expansions (cell-cell
interactions) and those which use only multipole expansions (cell-body
interactions with $\Delta=0$).  The scaling of these methods with $N$
depends on precisely how the error is constrained while increasing
$N$, but generally methods which support cell-cell interactions scale
as $O(N)$ and those that do not scale as $O(N \log N)$.  Our
experience has been that using $O(N)$-type algorithms for cosmological
simulation exposes some undesirable behaviors.  In particular, the
behavior of the errors near the outer regions of local expansions are
highly correlated.  To suppress the accumulation of these errors, the
accuracy of the local expansion must be increased, or their spatial
scale reduced to the point where the benefit of the $O(N)$ method is
questionable, at least at the modest accuracies of current
cosmological simulations.  For this reason, we have focused on
the implementation and optimization of an $O(N \log N)$ method.

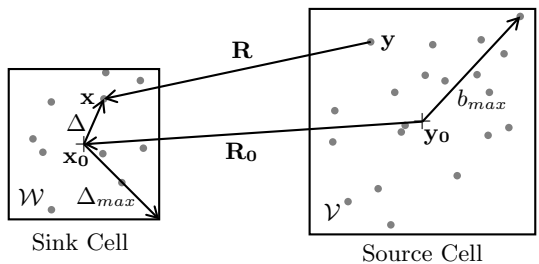
\begin{figure}[hbt]
\centering
\begin{tikzpicture}[>=angle 60]
\foreach \c in {
(1.251,1.074),
(0.454,1.093),
(0.322,1.271),
(1.290,2.153),
(1.265,1.803),
(1.504,0.689),
(0.566,1.762),
(0.568,0.328),
(1.790,1.147),
(1.750,2.047)
} \fill [black!50!white] \c circle (0.05);
\foreach \c in {
(5.800,2.043),
(4.335,1.646),
(6.373,1.318),
(4.308,1.230),
(4.812,2.561),
(5.080,0.127),
(6.226,2.125),
(6.8,2.9),
(4.911,0.608),
(5.637,2.524),
(5.219,1.362),
(5.270,1.451),
(4.511,0.436),
(5.962,0.781),
(6.649,1.421),
(5.537,2.117),
(5.132,1.890),
(6.390,1.890),
(6.585,2.403),
(6.281,2.589)
} \fill [black!50!white] \c circle (0.05);
\draw [thick] (0,0.2) rectangle (2,2.2);
\draw [thick] (4,0) rectangle (7,3);
\draw (1.0,1.2) node {+};
\draw (0.9,1.0) node {$\mathbf{x_0}$};
\draw (1.05,1.8) node {$\mathbf{x}$};
\draw (5.5,1.5) node {+};
\draw (5.7,1.3) node {$\mathbf{y_0}$};
\draw (5.05,2.55) node {$\mathbf{y}$};
\draw [thick,->] (4.812,2.561) -- (1.265,1.803);
\draw [thick,->] (5.5,1.5) -- (1.0,1.2);
\draw [thick,->] (1.0,1.2) -- (1.265,1.803);
\draw [thick,->] (1.0,1.2) -- (2.0,0.2);
\draw [thick,->] (5.5,1.5) -- (6.8,2.9);
\draw (0.9,1.48) node {$\Delta$};
\draw (1.3,0.5) node {$\Delta_{max}$};
\draw (6.3,1.8) node {$b_{max}$};
\draw (3.1,2.45) node {$\mathbf{R}$};
\draw (3.1,1.1) node {$\mathbf{R_0}$};
\draw (0.3,0.5) node {$\sW$};
\draw (4.3,0.3) node {$\sV$};
\draw (0.95,-0.1) node {Sink Cell};
\draw (5.5,-0.25) node {Source Cell};
\end{tikzpicture}
\caption{An illustration of the relevant distances used in the
multipole expansion and error bound equations.
\label{fig:mac}}
\end{figure}

Consider a configuration of sources as in Figure~\ref{fig:mac}.
The sources are contained within a ``source'' cell, $\sV$ of 
radius $b_{max}$, while
the field is evaluated at separation $\vD$ from $\vx_0$, the center 
of ``sink'' cell $\sW$.

In terms of an arbitrary Green's function, $G$, the field is:
\bea
\phi(\vx) &=& \int_\sV d\vy G(\vx-\vy) \rho(\vy) \label{eq:green}
\eea

Expanding G around $\vR_0 = \vx_0 - \vy_0$ in a Taylor series leads to
the Cartesian multipole expansion:
\be
\begin{split}
\phi(\vx) = \sum_{n=0}^p {(-1)^n \over n!} 
        \pnG (\vR_0) \odot \Mn \\ (\vy_0 + \vD)
        + \Phi_{(p)}(\vx)
\end{split}
\ee
where $\Phi_{(p)}$ is the error term, and 
the moment tensor is defined relative to a 
center, $\vz$ as:
\be
M^{(n)}(\vz) = 
        \int d\vy
          (\vy - \vz)^{(n)}  \rho(\vy)
\ee
We have used a notational shorthand in which
$\vv^{(n)}$ indicates
the n-fold outer product of the vector $\vv$ with itself,
while $\odot$ indicates a tensor inner-product
and $\pnG$ indicates the rank-$n$ tensor whose components are
the partial derivatives of G in the Cartesian directions.
We can further expand the result by writing 
$\Mn(\vy_0 + \vD)$ as a sum over powers of the components of $\vD$, and
then recollecting terms (see Eqns 12-14 in \cite{warren95a}).

\usepgflibrary{arrows}

While the mathematical notation above is compact, translating this
representation to an optimized interaction routine is non-trivial.
The expression for the force with $p=8$ in three dimensions begins
with $3^8 = 6561$ terms.  We resort to metaprogramming, translating the
intermediate representation of the computer algebra system
\cite{wolfram99} directly into \verb-C- code.  This approach is
capable of producing the necessary interaction routines through $p=8$
without human intervention.  A better approach would combine a
compiler with knowledge of the computing architecture into the
symbolic algebra system, allowing very high-level optimizations using
mathematical equivalences that are lost once the formulae are
expressed in a general programming language.  To our knowledge, no
such system currently exists.

We have also investigated support for pseudo-particle
\cite{kawai01} and kernel-independent \cite{ying04}
approaches which 
abstract the multipole interactions to more easily computed equations.  For
instance, the pseudo-particle method allows one to represent the far
field of many particles as a set of
pseudo-particle monopole interactions. We
have found that such approaches are not as efficient as a well-coded
multipole interaction routine in the case of gravitational or
Coulombic interactions, at least up to order $p=8$.


\subsection{Time Integration}

The original version of HOT integrated the equations of motion using
the leapfrog techniques described in \cite{efstathiou85}, with a
logarithmic timestep at early times.  This approach has proven
inadequate for high-accuracy simulations. Fortunately, the theory for
symplectic time integration in a comoving background was developed
by~\cite{quinn97}, which we have fully adopted.  The advantages of
this integrator are discussed in detail in \cite{springel05}.  We
calculate the necessary integrals for the ``drift'' and ``kick''
operators in arbitrary cosmologies with code added to the background
calculations in CLASS~\cite{lesgourgues11}.  We additionally restrict
the changes of the timestep to exact factors of two, rather than
allowing incremental changes at early times.  Any change of timestep
breaks the symplectic property of the integrator, but making
occasional larger adjustments rather than continuous small adjustment
(as is done in GADGET2 \cite{springel05}) appears to provide slightly better
convergence properties.  We have also modified 2HOT to save
``checkpoint'' files which maintain the leapfrog offset between
position and velocity.  This allows the code to maintain 2nd-order
accuracy in the time integration when restarting from a saved file.
Otherwise, the initial (first order) step in the leapfrog scheme can lead
to detectable errors after restarting at early times.

\subsection{Boundary Conditions}

Periodic boundary conditions have been applied to multipole methods
in a variety of ways, but most often are variants of the Ewald
method~\cite{hernquist91}.  For 2HOT, we have adopted the approach
described in~\cite{challacombe97}, which is based on the central
result of Nijboer \& De~Wette~(1957)~\cite{nijboer57}.  Effectively the same method in a
Cartesian basis was first used in a cosmological simulation
by~Metchnik~\cite{metchnik2009fast}.  This method sums the infinite series of
each relevant combination of powers of the co-ordinates, which can be
taken outside the sum of periodic replicas (since the multipole
expansion of each replica is identical).  These pre-computed
coefficients are then used in a local expansion about the center of
the volume.  We use $p=8$ and $ws=2$, which accounts for the boundary
effects to near single-precision floating point accuracy (one part in
$10^{-7}$).  The computational expense of this approach is about 1\%
of the total force calculation for the local expansion, and 5-10\% for
the 124 boundary cubes, depending on the overall accuracy tolerance.

\subsection{Force Smoothing}

The standard practice in cosmological N-body simulations is to smooth
the forces at small scales, usually with a Plummer or
spline~\cite{springel05} kernel.  We have implemented these smoothing
kernels in 2HOT, as well as the additional kernels described by
Dehnen~\cite{dehnen01}.  Dehnen concludes that the optimal softening
method uses a compensating kernel, with forces that are higher than
the Newtonian force at the outer edge of the smoothing kernel, which
compensates for the lower forces in the interior and serves to reduce
the bias in the force calculation.  Our tests confirm these
conclusions, and we use Dehnen's $K1$ compensating kernel for our
simulations, except for the tests comparing directly to other codes.

\section{Computational Approach}

\subsection{Domain Decomposition}

\begin{figure}[hbt]
\centering
\resizebox{1.0\columnwidth}{!}{\includegraphics{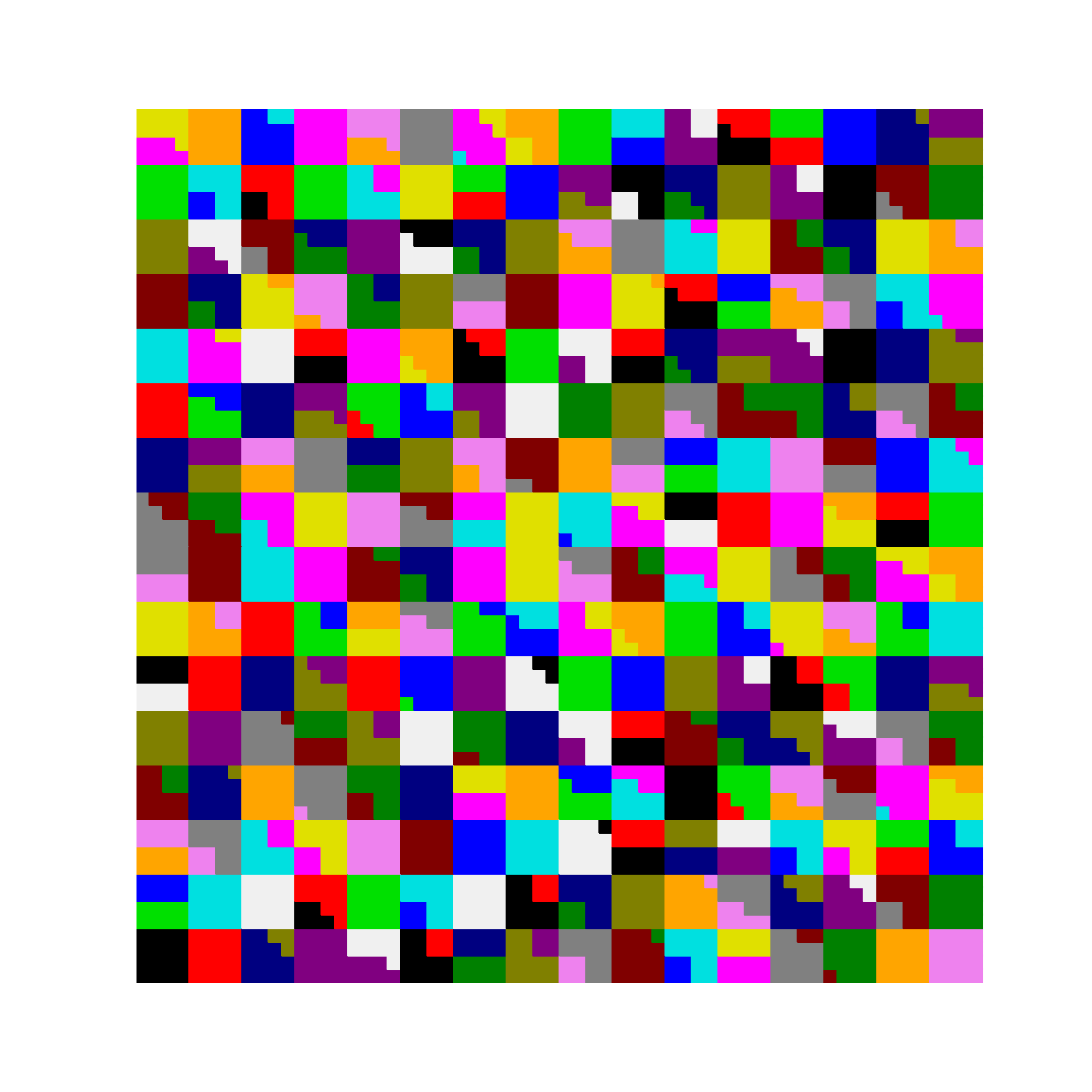}}
\caption{A demonstration of our space-filling curve domain
  decomposition for a highly evolved cosmological simulation on 3072
  processors in a cube 1 Gpc/h across.  We view one face of the 3-d
  computational volume, cycling through 16 different colors in turn
  for each processor domain. Starting in the lower left, the sequence
  goes black, red, green, blue, cyan, and then continues underneath in
  the $z$-dimension (not visible), returning to the front face with
  dark blue, brown, purple, white, etc.}
\label{fig:morton}
\end{figure}

The \emph{space-filling curve domain decomposition} approach we proposed
in WS93 has been widely adopted in both application codes
(e.g. \cite{griebel99,springel05,jetley08,wu12}) and more general
libraries~\cite{parashar96,macneice00}.  Our claim that such orderings
are also beneficial for improving memory hierarchy performance has
also been validated~\cite{springel05,mellor-crummey99}.  We show an
example of a 3-d decomposition of 3072 processor domains in Figure~\ref{fig:morton}.

The mapping of spatial co-ordinates to integer keys described in WS93
converts the domain decomposition problem into a generalized parallel
sort.  The method we use is similar to the sample sort described
in~\cite{solomonik10}, with the on-node portion done with an American flag radix
sort~\cite{mcllroy93}.  After using the samples to determine the edges of the
processor domains, in the initial HOT implementation the data was
moved using a loop over all pairs of processors needing to exchange
data. We converted the data exchange to use \verb-MPI_Alltoall()- for
improved scalability.  This exposed problems in the implementation of
Alltoall on large machines for both OpenMPI and the Cray system
MPI.  The first ``scalability surprise'' was related to the way
buffers were managed internally in OpenMPI, with the number of
communication buffers scaling as the number of processes squared.
This did not allow our code to run on more than 256 24-core nodes
using OpenMPI. We had to rewrite the implementation of Alltoall using
a hierarchical approach, with only one process per node relaying
messages to other nodes.  The second was a ``performance surprise'' as
defined by~\cite{thakur10}, where replacing the Cray system
implementation of \verb-MPI_Alltoall()- with a trivial implementation
using a loop over all pairs of processes exchanging data led to a huge
performance improvement when using more than 32k processors.  Note
that after the initial decomposition, the Alltoall communication
pattern is very sparse, since particles will only move to a small
number of neighboring domains during a timestep.  This also allows
significant optimization of the sample sort, since the samples can be
well-placed with respect to the splits in the previous decomposition.

\subsection{Tree Construction and Traversal}

The parallel tree construction in WS93 used a global concatenation of
a set of ``branch'' nodes from each processor to construct the tree at
levels coarser than the individual processor domains.  While this is
an adequate solution up to a few thousand processors, at the level of
tens of thousands of domains and larger, it leads to unacceptable
overhead.  Most of the nodes communicated and stored will never be
used directly, since the local traversal will only probe that deeply
in the tree near its own spatial domain.  Instead of a global
concatenation, we proceed with a pairwise hierarchical aggregation
loop over $i$ up to $log_2\thinspace N_{proc}$ by exchanging branch
nodes between nearest neighbors in the 1-d space-filling curve,
incrementally updating the tree with those nodes, then doing the same
with the $2^i$-th neighbor.  This provides a minimal set of shared
upper-level nodes for each processor domain, and has demonstrated its
scalability to 256k processors.

In \cite{warren95a} we describe a tree traversal abstraction which
enables a variety of interactions to be expressed between ``source''
and ``sink'' nodes in tree data structures.  This abstraction has
since been termed \emph{dual-tree traversal}~\cite{gray01,yokota12}.
The dual-tree traversal is a key component of our method to increase
the instruction-level parallelism in the code to better enable new CPU and GPU
architectures (see Section~\ref{sec:ilp}).  

During the tree traversal we use the same request/reply protocol
described in WS93 using the global key labels assigned during the tree
construction phase.  Additional bits to label the source processor
have been added to the hcells to support machines with up to $2^{18}$
processors.  Our initial approach to hiding latency in the tree
traversal was recast in the form of an active message abstraction.  We
believe that such event-driven handlers are more robust and less
error-prone to implement correctly~\cite{ousterhout96}.  We currently
use our own implementation of active messages within MPI, which we
call ``Asynchronous Batched Messages'' (ABM). ABM is a key component
of our ability to overlap communication and computation and hide
message latency.  MPI has supported one-sided communications
primitives for many years, but their performance is often worse than
regular point-to-point communication.  It is likely that
synchronization and locking overheads and complexity are to
blame~\cite{balaji11}.  Newer implementations of active
messages~\cite{willcock10} are an attractive alternative, which we
plan to implement as time allows.

\subsection{Improving instruction-level parallelism}
\label{sec:ilp}

In WS93 we used the fact that particles which are spatially near each
other tend to have very similar cell interaction lists.  By updating
the particles in an order which takes advantage of their spatial
proximity, we improved the performance of the memory hierarchy.  Going
beyond this optimization with dual-tree traversal, we can bundle a set
of $m$ source cells which have interactions in common with a set of
$n$ sink particles (contained within a sink cell), and perform the
full $m \times n$ interactions on this block.  This further improves
cache behavior on CPU architectures, and enables a simple way for GPU
co-processors to provide reasonable speedup, even in the face of
limited peripheral bus bandwidth.  We can further perform data
reorganization on the source cells (such as swizzling from an
array-of-structures to a structure-of-arrays for SIMD processors) to
improve performance, and have this cost shared among the $n$ sinks.
In an $m \times n$ interaction scheme, the interaction vector for a
single sink is computed in several stages, which requires writing the
intermediate results back to memory multiple times, in contrast to the
WS93 method which required only one write per sink.  For current
architectures, the write bandwidth available is easily sufficient to
support the $m \times n$ blocking.

Taking advantage of instruction-level parallelism is essential.  In
the past, obtaining good CPU performance for gravitational kernels
often required hand-tuned assembly code.  Implementing the complex
high-order multipole interactions using assembly code would be
extremely difficult.  Fortunately, the gcc compiler comes to the
rescue with vector intrinsics \cite{stallman89}.  We use gcc's
\verb-vector_size- attribute, which directs the compiler to use
\verb-SSE- or \verb-AVX- vector instructions for the labeled
variables.  By providing the interaction functions with the
appropriately aligned and interleaved data, gcc is able to obtain near
optimal SIMD performance from C code.

We have also implemented our gravitational interaction functions with
both CUDA and OpenCL kernels on NVIDIA GPUs, obtaining single-precision
performance of over 2 Tflops on a K20x (Table~\ref{tab:ukernel}).  We have implemented
these kernels within 2HOT and demonstrated a 3x speedup over using the
CPU alone.  The ultimate performance of our code on hybrid GPU
architectures depends on the ability of the to perform a highly
irregular tree-traversal quickly enough to provide the necessary flow
of floating-point intensive gravitational interactions.  A parallel
scan and sort based on our space-filling curve key assignment is one
example of a successful approach \cite{bedorf12}.

We have generally achieved near 40\% of peak (single-precision) CPU
performance on the supercomputers we have ported our code to over the
past 20 years.  We are working toward demonstrating the performance of
2HOT on Titan, using 18,688 NVIDIA K20x GPUs.  With 25\% of peak
performance, we would obtain near 20 Tflops on that machine.

\subsection{Managing the Simulation Pipeline}

In order to better integrate the various codes involved, and to
simplify the management of the multiple configuration files per
simulation, we have developed a Python \cite{vanrossum95}
metaprogramming environment to translate a high-level description of a
simulation into the specific text configuration files and shell
scripts required to execute the entire simulation pipeline.  Without
this environment, it would be extremely difficult to guarantee
consistency among the various components, or to reproduce earlier
simulations after new features have been added to the individual
software agents.  It also allows us to programatically generate the
configuration of thousands of simulations at once, that would
previously have to be configured manually.

\subsubsection{Task Management}
Modern simulation pipelines present a complex task for queueing
systems.  Given the flexibility of 2HOT, which can run on an arbitrary
number of processors, or be interrupted with enough notice to write a
checkpoint, we would like to control our tasks using higher-level
concepts.  We wish to specify the general constraints on a simulation
task and have the system perform it in an efficient manner with as
little human attention as possible.  For example, ``Please run our
simulation that will require 1 million core-hours using as many jobs
in sequence as necessary on at least 10,000 cores at a time, but use
up to 2x as many cores if the wait for them to become available does
not increase the overall wallclock time, and allow our job to be
pre-empted by higher-priority jobs by sending a signal at least 600
seconds in advance.''  Optimal scheduling of such requests from
hundreds of users on a machine with hundreds of thousands of
processors is NP-hard, but there seems to be ample room for
improvement over the current systems, even without an ``optimal''
solution.

Data analysis often requires many smaller tasks, which queueing
systems and MPI libraries have limited support for as well.  We have
developed an additional Python tool called \verb-stask-.  It allows us
to maintain a queue inside a larger PBS or Moab allocation which can
perform multiple smaller simulations or data analysis tasks.  It has
also proven useful to manage tens of thousands of independent tasks
for \verb-MapReduce- style jobs on HPC hardware.  For instance, we
have used this approach to generate 6-dimensional grids of
cosmological power spectra, as well as perform Markov-Chain Monte
Carlo analyses.

\subsubsection{Checkpoints and I/O}
2HOT reads and writes single files using collective MPI/IO routines.
We use our own self-describing file format (SDF), which consists of
ASCII metadata describing raw binary particle data structures.  I/O
requirements are driven primarily by the frequency of checkpoints,
which is in turn set by the probability of failure during a run.  For
the production simulations described here, we experience a hardware
failure which ends the job about every million CPU hours (80 wallclock
hours on 12288 CPUs).  Writing a 69 billion particle file takes about
6 minutes, so checkpointing every 4 hours with an expected failure
every 80 hours costs 2 hours in I/O and saves 4-8 hours of
re-computation from the last permanently saved snapshot.  At LANL, we
typically obtain 5-10 Gbytes/sec on a Panasas filesystem.  We have
demonstrated the ability to read and write in excess of 20 Gbytes/sec
across 160 Lustre OSTs on the filesystem at ORNL.  By modifying our
internal I/O abstraction to use MPI/IO across 4 separate files to
bypass the Lustre OST limits, we have obtained I/O rates of 45
Gbytes/sec across 512 OSTs.  These rates are sufficient to support
simulations at the $10^{12}$ particle scale at ORNL, assuming the
failure rate is not excessive.

\subsubsection{Version Control of Source Code and Data}

To assure strict reproducibility of the code and scripts used for any
simulation and to better manage development distributed among multiple
supercomputer centers, we use the \verb-git- version control
system~\cite{torvalds05} for all of the codes in the simulation
pipeline, as well as our Python configuration system.  We additionally
automatically propagate the \verb-git- tags into the metadata included
in the headers of the data which is produced from the tagged software.

\subsubsection{Generating Initial Conditions}

We use the Boltzmann code CLASS~\cite{lesgourgues11,blas11} to
calculate the power spectrum of density fluctuations for a particular
cosmological model.  A particular realization of this power spectrum
is constructed using a version of 2LPTIC~\cite{crocce06} we
have modified to support more than $2^{31}$ particles and use the FFTW3
library.

\subsubsection{Data Analysis}

One of the most important analysis tasks is generating halo catalogs
from the particle data by identifying and labeling groups of
particles.  We use \verb-vfind-~\cite{pfitzner98} implemented with
the HOT library to perform both friend-of-friends (FOF) and isodensity
halo finding.  More recently, we have adopted the ROCKSTAR halo
finder~\cite{behroozi13}, contributing some scalability
enhancements to that software, as well as interfacing it with SDF.
Our plans for future data analysis involve developing interfaces
to the widely-adopted \verb-yt- Project~\cite{turk11}, as well as
contributing the parallel domain decomposition and tree traversal technology 
described here to \verb-yt-.

Many of the mathematical routines we developed over the years as needed for our
evolution or analysis codes have been replaced with superior
implementations.  The GSL~\cite{galassi07} and FFTW~\cite{frigo98}
libraries have been particularly useful.

\section{Scalability and Performance}
In Table~\ref{tab:nbody-historical} we show the performance of our N-body code on a sample of the
major supercomputer architectures of the past two decades.  It is
perhaps interesting to note that now a single core has more memory and
floating-point performance than the fastest computer in the world in
1992 (the Intel Delta, on which we won our first Gordon Bell prize~\cite{warren92b}).
We show a typical breakdown among different phases of our code in Table~\ref{tab:phases}, and
single processor performance in Table~\ref{tab:ukernel}.

\begin{table}[htb]
\centering
\begin{tabular}{|c|c|c|r|r|r|r|}
\hline
Year & Site & Machine & Procs & Tflop/s \\
\hline\hline
2012 & OLCF & Cray XT5 (Jaguar) & 262144 & 1790 \\
\hline
2012 & LANL & Appro (Mustang) & 24576 & 163 \\
\hline
2011 & LANL & SGI XE1300 & 4096 &  41.7 \\ 
\hline
2006 & LANL & Linux Networx& 448 & 1.88 \\ 
\hline
2003 & LANL  & HP/Compaq (QB) & 3600 & 2.79 \\
\hline
2002 & NERSC & IBM SP-3(375/W) & 256 & 0.058 \\
\hline
1996 & Sandia  & Intel (ASCI Red) & 6800 &  0.465 \\
\hline
1995 & JPL     & Cray T3D      & 256 &     0.008 \\
\hline
1995 & LANL    & TMC CM-5      & 512 &    0.014 \\
\hline
1993 & Caltech & Intel Delta   & 512 &    0.010 \\
\hline
\end{tabular}
\caption{Performance of HOT on a variety of parallel supercomputers spanning
20 years of time and five decades of performance.}
\label{tab:nbody-historical}
\end{table}


\begin{table}[htb]
\centering
\begin{tabular}{rr}
\hline
{\em computation stage} & {\em time} (sec)\\
\hline
Domain Decomposition & 12\\
Tree Build & 24\\
Tree Traversal & 212\\
Data Communication During Traversal & 26\\
Force Evaluation & 350\\
Load Imbalance & 80\\
\hline
Total (56.8 Tflops) & 704\\
\hline
\end{tabular}
\caption{Breakdown of computation stages in a single timestep from a recent $4096^3$
  particle simulation using 2HOT on 12288 processors of Mustang at
  LANL.  The force evaluation consisted of 1.05e15 hexadecapole interactions, 
  1.46e15 quadrupole interactions and 4.68e14 monopole interactions, for a total of
  582,000 floating point operations per particle.  Reducing the accuracy parameter
  to a value consistent with other methods would reduce the operation count by more than a
  factor of three.}
\label{tab:phases}
\end{table}

\begin{figure}[htb]
\centering
\includegraphics[width=3.25in]{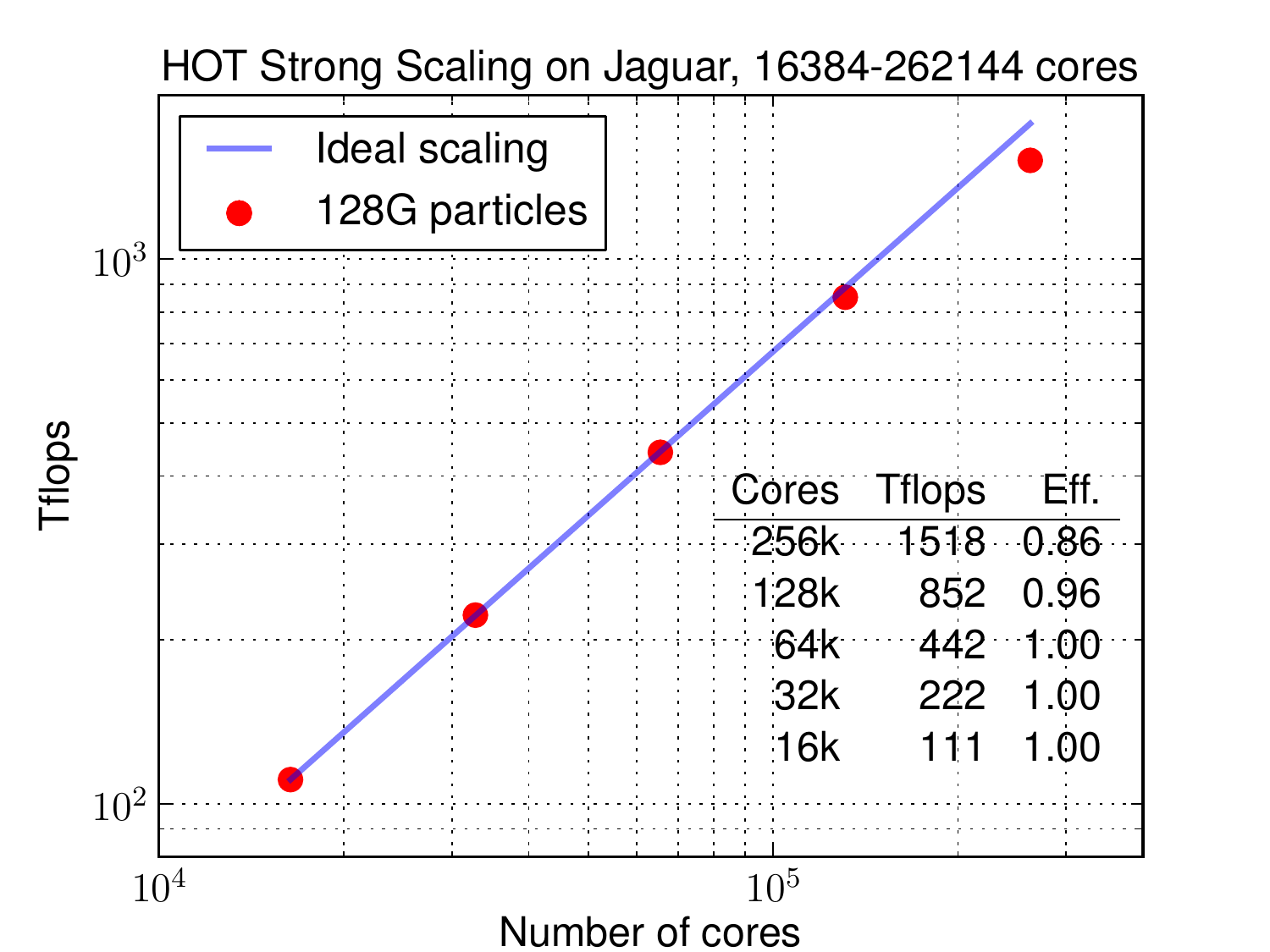}
\caption{Scaling on Jaguar measured in June 2012.}
\label{fig:scaling}
\end{figure}

We present strong scaling results measured on Jaguar in
Figure~\ref{fig:scaling}.  These benchmarks represent a single
timestep, but are representative of all aspects of a production
simulation, including domain decomposition, tree construction, tree
traversal, force calculation and time integration, but do not include
I/O (our development allocation was not sufficient to perform this set
of benchmarks if they had included I/O).  Also, note that these
results were using the code prior to the implementation of background
subtraction, so the error tolerance was set to a value resulting in
about 4 times as many interactions as the current version of the code
would require for this system.

\begin{table}[htb]
\centering
\begin{tabular}{|l|r|}
\hline  Processor                   & Gflop/s \\ \hline
\hline  
       2530-MHz Intel P4 (icc) &            1.17 \\ \hline
       2530-MHz Intel P4 (SSE) &            6.51 \\ \hline
       2600-MHz AMD Opteron 8435 &         13.88 \\ \hline
       2660-MHz Intel Xeon E5430 &         16.34 \\ \hline
       2100-MHz AMD Opteron 6172 (Hopper) & 14.25 \\ \hline 
       PowerXCell 8i (single SPE) &        16.36 \\ \hline
       2200-MHz AMD Opteron 6274  (Jaguar) & 16.97 \\ \hline
       2600-MHz Intel Xeon E5-2670 (AVX) & 28.41 \\ \hline
       1300-MHz NVIDIA M2090 GPU (16 SMs) & 1097.00 \\ \hline
       732-MHz NVIDIA K20X GPU (15 SMs) & 2243.00 \\ \hline
\end{tabular}
\caption{Single core/GPU performance in Gflop/s obtained with our gravitational micro-kernel benchmark for the monopole interaction.  All numbers are for single-precision calculations, calculated using 28 flops per interaction.}
\label{tab:ukernel}
\end{table}

\section{Error Analysis}

Verifying the correctness of a large simulation is a complex and
difficult process.  Analogous to the ``distance ladder'' in astronomy,
where no single technique can measure the distances at all scales
encountered in cosmology, we must use a variety of methods to check
the results of our calculations.  As an example, using the
straightforward Ewald summation method to calculate the force on a
single particle in a $4096^3$ simulation requires over $10^{14}$
floating point operations (potentially using 128-bit quadruple
precision), so it is impractical to use for more than a very small
sample of particles.  However, it can be used to verify a faster
method, and the faster method can be used to check the accuracy of the
forces in a much larger system.  Eventually, we reach the stage where
we can use 2HOT itself to check lower-accuracy results by adjusting
the accuracy parameter within the code (as long
as we are willing to pay the extra cost in computer time for higher
accuracy).

\begin{figure}[htb]
\centering
\resizebox{0.9\columnwidth}{!}{\includegraphics{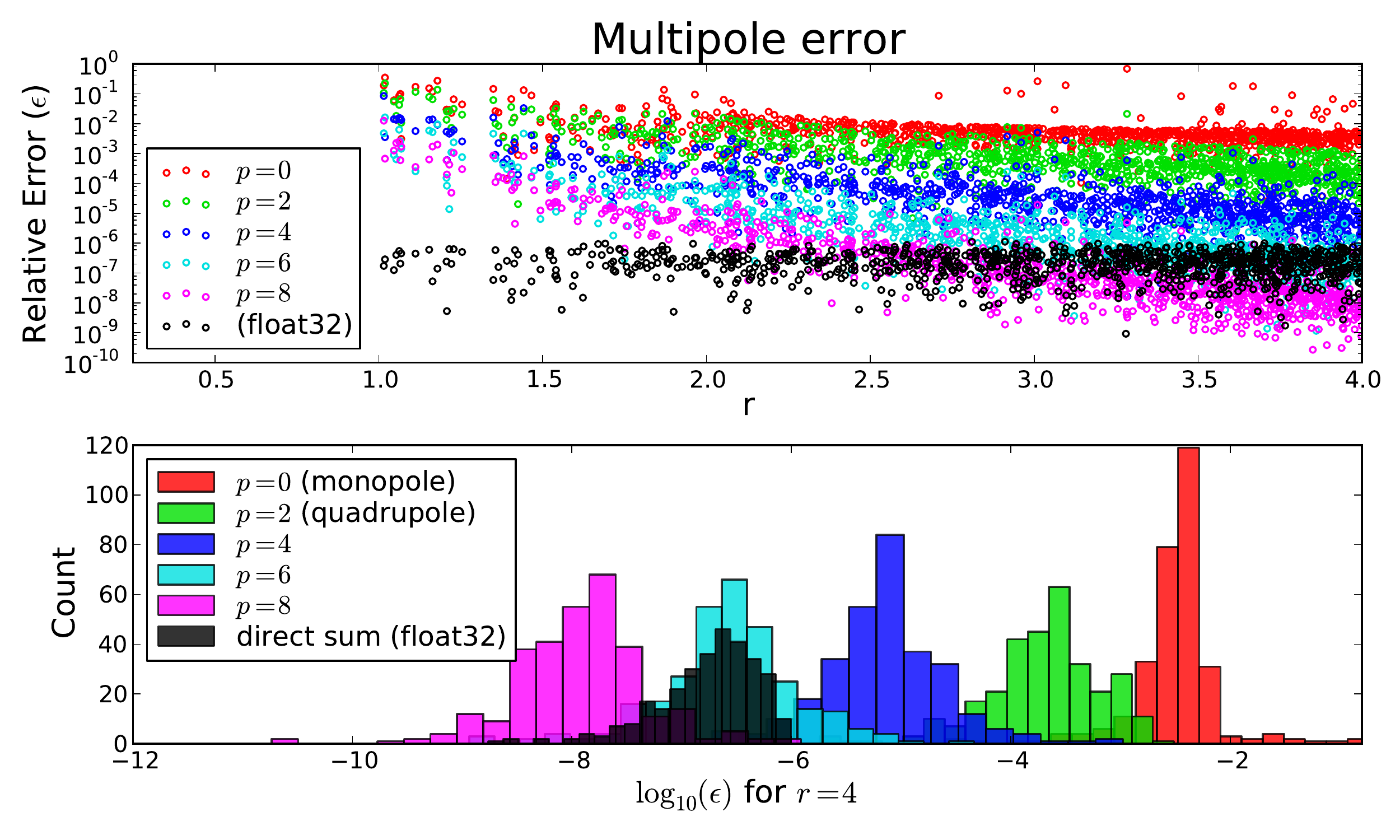}}
\caption{Error behavior for multipoles of various order ($p$) for 512
  particles randomly distributed in a cube of size 1 at distance
  $r$. A single $p=8$ multipole is more accurate than direct summation
  in single precision at $r=4$.}
\label{fig:p8err}
\end{figure}

Additionally, writing simple tests to verify the behavior of
individual functions is essential.  We have used
Cython~\cite{behnel11} to wrap the functions in 2HOT, allowing them to
be tested from within a more flexible and efficient Python
environment.  In Figure~\ref{fig:p8err} we show one such example,
showing the expected behavior of various orders of multipole
interactions vs distance.

We also can compare the results of 2HOT with other codes, and
investigate the convergence properties of various parameters.  One
must always keep in mind that convergence testing is necessary, but
not sufficient, to prove correctness.  In a complex system there may
be hidden parameters that are not controlled for, or variables that
interact in an unexpected way, reducing the value of such tests.
Having two methods agree also does not prove that they are correct,
only that they are consistent.

\begin{figure*}[tb!]
\centering
\resizebox{1.6\columnwidth}{!}{\includegraphics{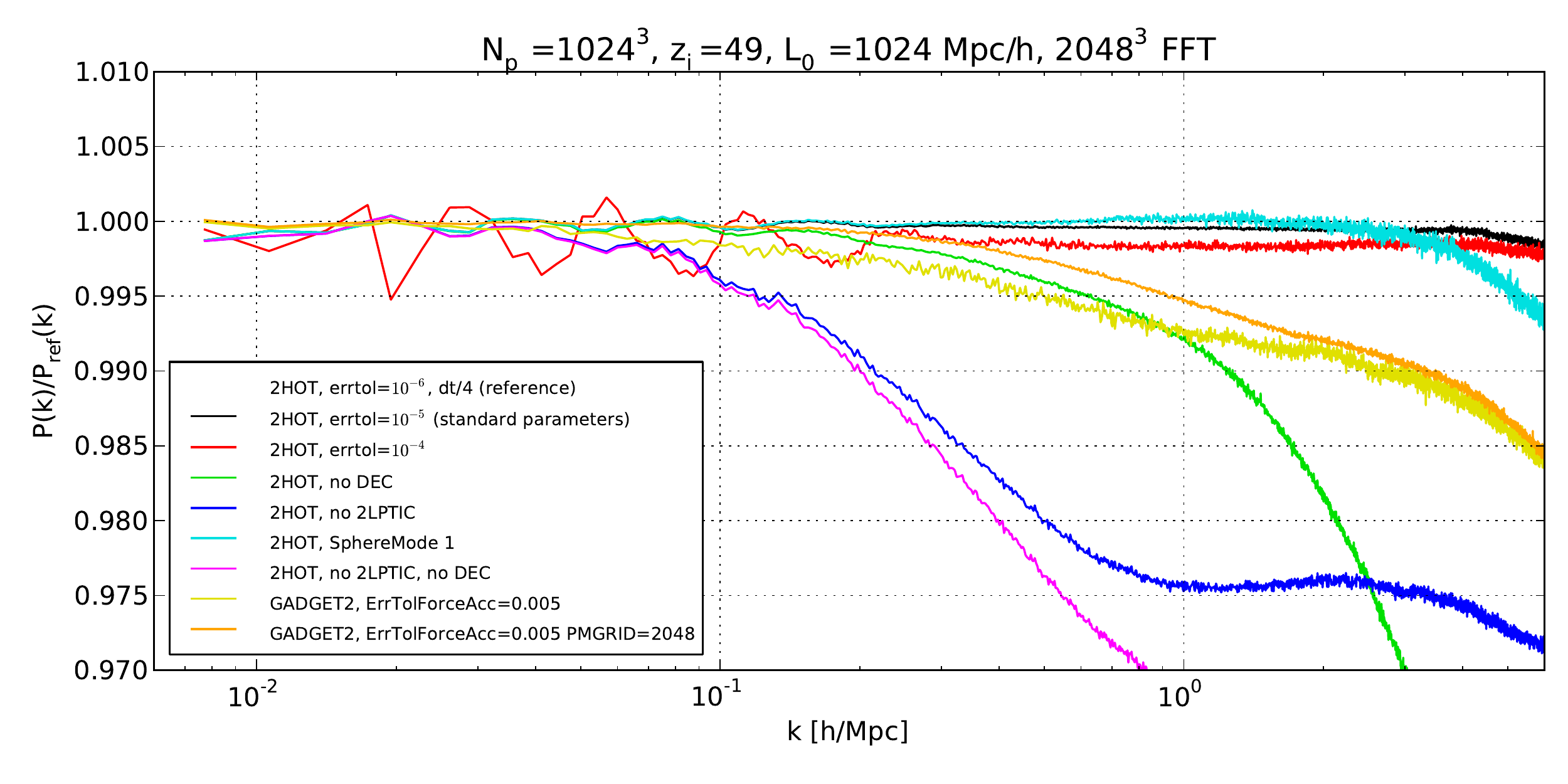}}
\resizebox{1.6\columnwidth}{!}{\includegraphics{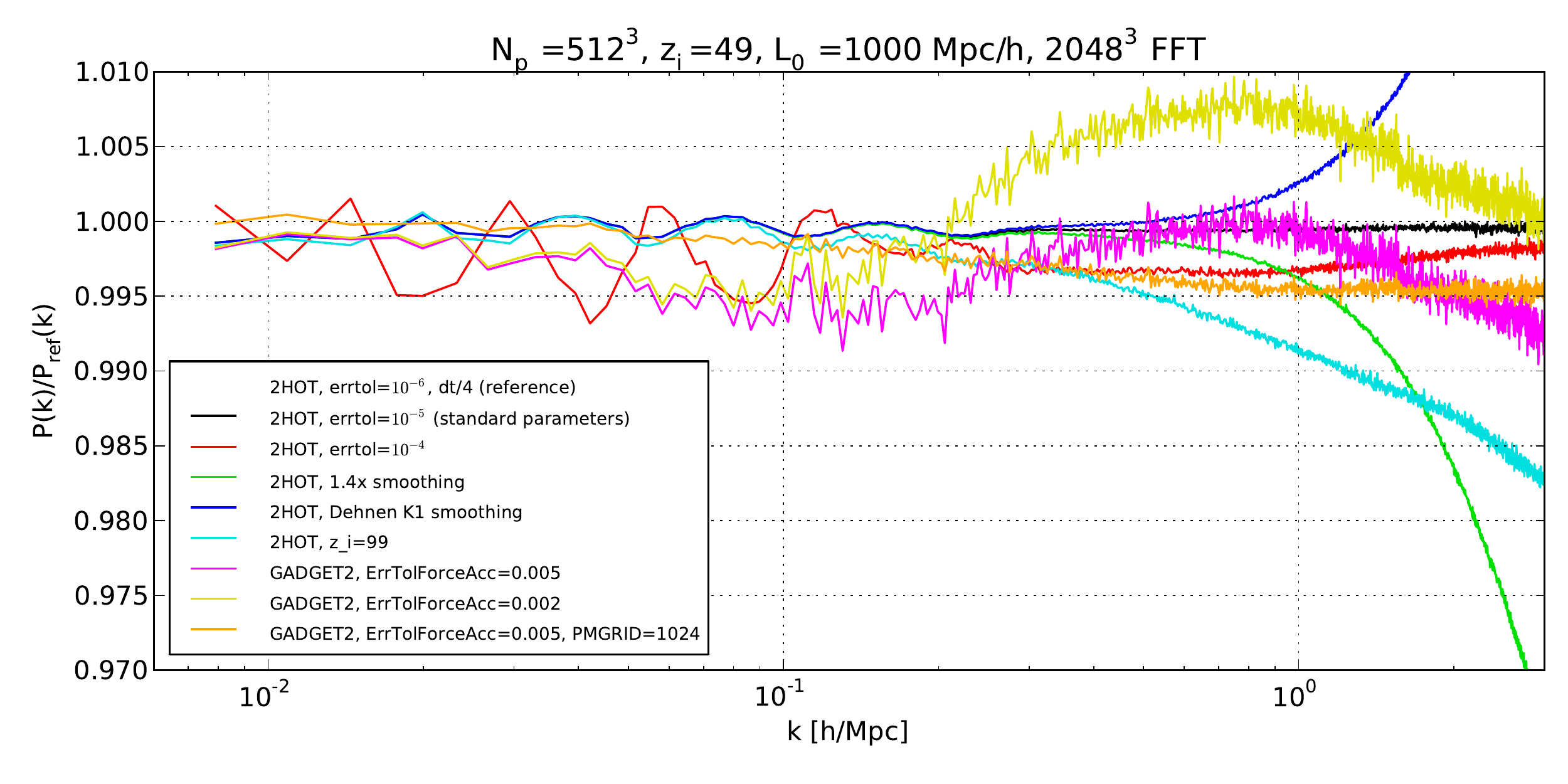}}
\caption{We show the difference between the power spectra at z=0 using 2HOT
  and GADGET2 on $1024^3$ particles in a 1Gpc/h box, as well as
  variations due to using the 2LPT correction to the initial
  conditions and discretization error correction (DEC) of the
  same form as a cloud-in-cell deconvolution..
  GADGET2 and 2HOT agree within 0.1\% on large spatial scales, but
  GADGET2 is about 1\% lower at k=1, in its TreePM transition region.
  With a relaxed error tolerance of $10^{-4}$ (10x the value used for
  our scientific results, which results in a 3x reduction in the
  number of interactions) 2HOT demonstrates errors at the 0.5\% level
  at large scales, with errors at the 0.1\% level at small scales.
  Not using 2LPTIC initial conditions reduces the power spectrum at
  k=1 by more than 2\% (blue curve). The lower panel shows the same graphs
  using a lower $512^3$ particle resolution (8x higher particle mass).
  We note the GADGET2 results differ among themselves at the 0.5\%
  level, depending on the chosen parameters.  The change in resolution
  moves the TreePM transition region for GADGET2 to a spatial scale 2x
  as large (k reduced a factor of 2) compared with the previous
  figure.  The effects of changing the smoothing kernel (blue) and smoothing
  length (green) are also shown.}
\label{fig:pspech_10_cps2}
\end{figure*}

In Figure~\ref{fig:pspech_10_cps2} we show the sensitivity of the
power spectrum to adjustments in various code parameters, as well as
comparing with the widely used GADGET2~\cite{springel05} code.  The
power spectrum is a sensitive diagnostic of errors at all spatial
scales, and can detect deficiencies in both the time integration and
force accuracy.  We can conclude from these graphs that 2HOT with the
settings used for our scientific results (an error tolerance of
$10^{-5}$) produces power spectra accurate to 1 part in 1000 at
intermediate and large scales, with parameters such as the smoothing
length and starting redshift dominating over the force errors at small
scales.  2HOT also systematically differs from GADGET2 at scales
corresponding to the switch between tree and particle-mesh, an effect
also observed when comparing GADGET2 with perturbation theory
results at high redshift~\cite{taruya12}.

\section{Scientific Results}
\label{sec:science}

The number of objects in the Universe of a given mass is a fundamental
statistic called the mass function.  The mass function is sensitive to
cosmological parameters such as the matter density, $\Omega_m$, the
initial power spectrum of density fluctuations, and the dark energy
equation of state.  Especially for very massive clusters (above
$10^{15}$ solar masses [$M_\odot/h$]) the mass function is a sensitive probe of
cosmology.  For these reasons, the mass function is a major target of
current observational programs \cite{planckcollaboration13a}.  Precisely
modeling the mass function at these scales is an enormous challenge
for numerical simulations, since both statistical and systematic
errors conspire to prevent the emergence of an accurate theoretical
model (see \cite{reed12} and references therein).  The dynamic range
in mass and convergence tests necessary to model systematic errors
require multiple simulations at different resolutions, since even a
$10^{12}$ particle simulation does not have sufficient
statistical power by itself.

\begin{figure}[hbt]
\centering
\includegraphics[width=1.05\columnwidth]{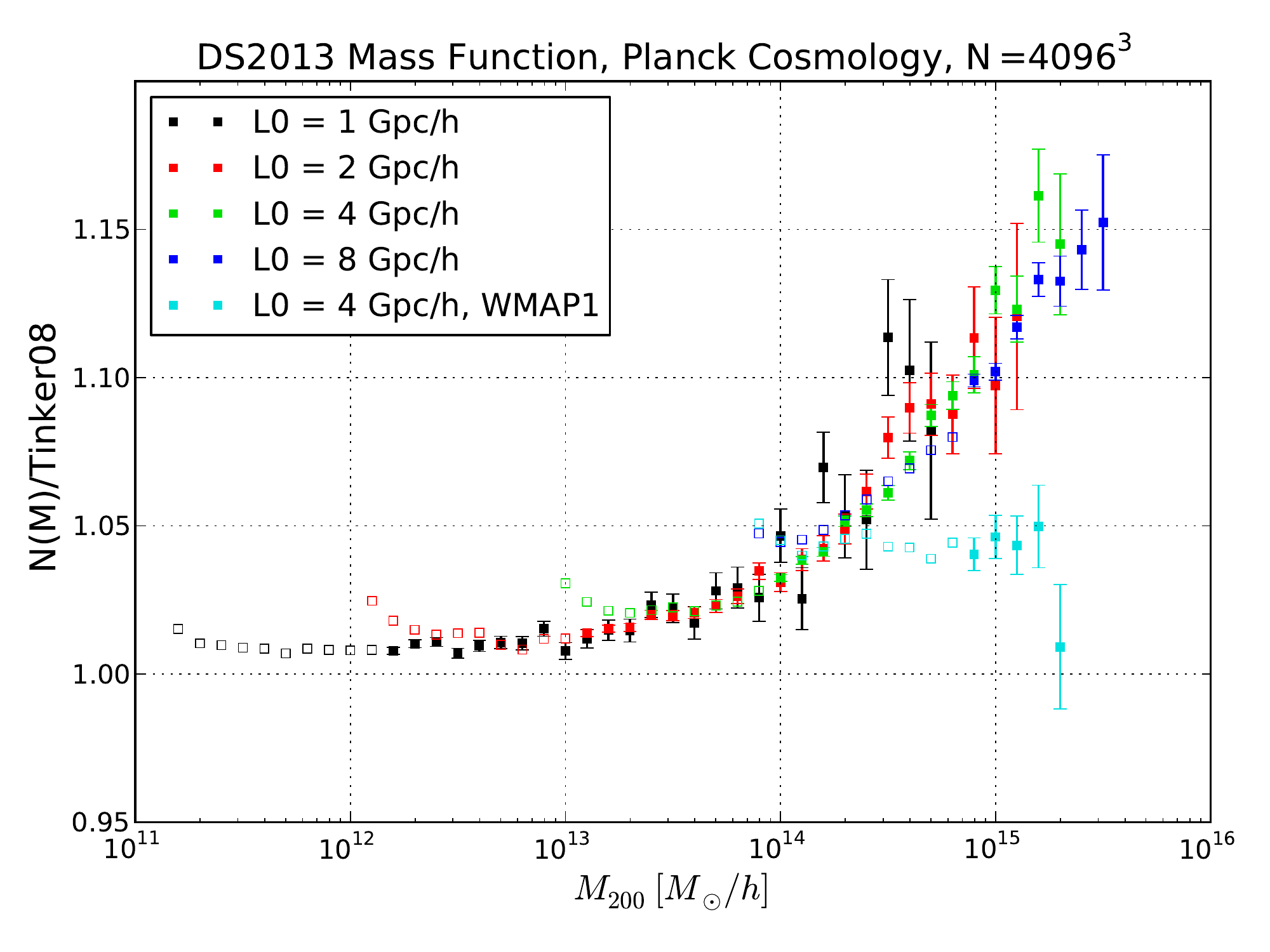}
\caption{A plot of the mass function from four recent $4096^3$
  particle simulations computed with 2HOT.  The scale of the
  computational volume changes by a factor of two between each
  simulation (so the particle mass changes by factors of 8).  We plot
  our data divided by the fit of
  Tinker08~\protect\cite{tinker08} on a linear $y$-axis.  The figure
  shows the simulations are internally consistent 
  but deviate from the Tinker08 fit at large
  scales.  Open symbols are used for halos with 100-1000 particles, showing
  consistency at the 1\% level down to 200 particles per halo.}
\label{fig:ds2013mf}
\end{figure}

Our HOT code was an instrumental part of the first calculations to
constrain the mass function at the 10\% level \cite{warren06} with a
series of sixteen $1024^3$ simulations performed in 2005, accounting
for about $4 \times 10^{18}$ floating point operations. These results
were further refined to a 5\% level of accuracy with the addition of
simulations from other codes, and the use of a more observationally
relevant spherical overdensity (SO) mass definition \cite{tinker08}.
With our suite of simulations (twelve $4096^3$
simulations, with an aggregate volume of thousands of cubic Gpc, using
roughly 20 million core-hours and accounting for $4 \times 10^{20}$
floating point operations), we are able to probe effects at the 1\%
level in the SO mass function above $10^{15} M_\odot/h$ for the first
time.

Some highlights of our scientific results for the mass function of dark
matter halos (Figure~\ref{fig:ds2013mf}) are:
\begin{itemize}
\item We provide the first mass function calculated from a suite of
  simulations using the new standard Planck 2013 cosmology (with a
  $4096^3$ particle simulation and six $2048^3$ simulations completed
  and shared with our collaborators within 30 days of the publication
  of the Planck 2013 results).  Changes in the parameters from the
  previous WMAP7 model are large enough that extrapolations from the
  other cosmologies~\cite{angulo10,angulo12} are likely subject to
  systematic errors which are large compared to the statistical
  precision of our results.
\item We find the Tinker08 \cite{tinker08} result underestimates the
  mass function at scales of $10^{15} M_\odot/h$ by about 5\% when
  compared with the older WMAP1 cosmological model it was calibrated
  against.
\item For the Planck 2013 cosmology, the Tinker08 mass function is
  10-15\% low at large scales, due to the added systematic effect of
  non-universality in the underlying theoretical model.
\item We identify a systematic error stemming from the improper growth
  of modes near the Nyquist frequency, due to the discrete
  representation of the continuous Fourier modes in the ideal input
  power spectrum with a fixed number of particles.  This is a
  resolution dependent effect which is most apparent when using
  particle masses larger than $10^{11} M_\odot$ (corresponding to
  using less than 1 particle per cubic Mpc/h).  Uncertainty in the
  appropriate correction and consequences of this effect appear to be
  the dominant source of systematic error in our results, where
  statistical uncertainties prevent us from ruling out a 1\%
  underestimate of the mass function at scales of $2 \times 10^{15}
  M_\odot/h$ and larger.  If uncontrolled, this discretization error
  confounds convergence tests which attempt to isolate the effects of
  the starting redshift of the simulation \cite{luki07,reed12}, since
  the error becomes larger at higher starting redshifts.
\item We are in direct conflict with recent results \cite{watson12}
  (see their Figure 13) which find the SO mass function to be lower
  than the the Tinker08 result at high masses.  Potential explanations
  would be insufficient force accuracy of the CUBEP$^3$M code
  \cite{harnois-deraps12} (c.f.~their Figure 7 showing force errors of
  order 50\% at a separation of a few mesh cells), with a secondary
  contribution from initial conditions that did not use
  2LPT~\cite{crocce06} corrections (more recent simulations
  in~\cite{watson13} appear consistent with our results up to $2
  \times 10^{15} M_\odot/h$).
\end{itemize}

\section{Conclusion}

Using the background subtraction technique described in
Section~\ref{sec:bs} improved the efficiency of our treecode algorithm
for cosmological simulations by about a factor of three when using a
relatively strict tolerance ($10^{-5}$), resulting in a total absolute
force error of about 0.1\% of the typical force.  We have evidence
that accuracy at this level is required for high-precision scientific
results, and we have used that tolerance for the results presented
here.  That accuracy requires about 600,000 floating point operations
per particle (coming mostly from $\sim$2000 hexadecapole
interactions).  Relaxing the error parameter by a factor of 10
(reducing the total absolute error by a factor of three) reduces the
operation count per particle to 200,000.

We can compare our computational efficiency with the 2012 Gordon Bell
Prize winning TreePM N-body application~\cite{ishiyama12} which used
140,000 floating point operations per particle.  The $\theta$
parameter for the Barnes-Hut algorithm in that work was not specified,
so it is difficult to estimate the effective force accuracy in their
simulation.  Modulo being able to precisely compare codes at the same
accuracy, this work demonstrates that a pure treecode can be
competitive with TreePM codes in large periodic cosmological volumes.
The advantage of pure treecodes grows significantly as applications
move to higher resolutions in smaller volumes, use simulations with
multiple hierarchical resolutions, and require non-periodic boundary
conditions.

Our experience with HOT over the past twenty years perhaps provides a
reasonable baseline to extrapolate for the next ten years.  The Intel
Delta machine provided 512 single processor nodes running at 40 MHz
and no instruction-level parallelism (concurrency of 512).  The
benchmark we ran on Jaguar had 16,384 16-core nodes running at 2.2GHz
and 4-wide single-precision multiply-add SSE instructions (concurrency
of 2.1 million).  The performance difference for HOT of 180,000
between these machines is nicely explained from a factor of 55 in
clock rate, a factor of 4096 in concurrency, and the loss of about
20\% in efficiency.  (Most of the efficiency loss is simply the fact
that the gravitational inner loop can not balance multiplies and adds,
so FMA instructions can not be fully utilized).

Looking to the future, if we guess clock rates go down a factor of two
for better power utilization, and we lose up to a factor of two in
efficiency, we would need an additional factor of 2000 in concurrency
to reach an exaflop.  A factor of 64 is gained going to 256-wide
vector operations, leaving us with 32x as many cores.  A machine with
8 million cores is daunting, but measured logarithmically the jump
from $\log_2(512) = 9$ on the Delta to $\log_2(262144) = 18$ on Jaguar
is twice as large as the jump from Jaguar to an exaflop machine with
$\log_2(N_{cores})$ of 23.  Assuming the hardware designers make
sufficient progress on power and fault-tolerance challenges, the basic
architecture of 2HOT should continue to serve at the exascale level.

\section{Acknowledgments}

We gratefully acknowledge John Salmon for his many contributions to
the initial version of HOT, and helpful comments on a draft version of
this manuscript.  We thank Mark Galassi for his memory management
improvements to 2HOT and Ben Bergen for assistance with the OpenCL
implementation.  We thank the Institutional Computing Program at LANL
for providing the computing resources used for our production
simulations.  This research used resources of the Oak Ridge Leadership
Computing Facility at Oak Ridge National Laboratory, which is
supported by the Office of Science of the Department of Energy under
Contract DE-AC05-00OR22725.  This research also used resources of the
National Energy Research Scientific Computing Center, which is
supported by the Office of Science of the U.S. Department of Energy
under Contract No. DE-AC02-05CH11231. This research was performed
under the auspices of the National Nuclear Security Administration of
the U.~S.~Department of Energy under Contract DE-AC52-06NA25396.

\bibliographystyle{msw2}
\bibliography{refs,../zotero}
\end{document}